\documentclass[11pt,twoside]{article}

\usepackage{asp2006} \usepackage{epsf} \usepackage{lscape}
\usepackage{graphicx}

\begin{document}
\title{Liverpool Telescope Optical Photometry Following the 2006
  Outburst of RS Ophiuchi} \author{Matthew~J.~Darnley,
  Rebekah~A.~Hounsell, and Michael~F.~Bode} \affil{Astrophysics
  Research Institute, Liverpool John Moores University, Twelve Quays
  House, Egerton Wharf, Birkenhead, CH41 1LD, UK}


\begin{abstract}
We present a preliminary report on the broadband optical photometry of
the 2006 outburst of the recurrent nova RS Ophiuchi.  These data were
obtained using the robotic 2m Liverpool Telescope and cover the
outburst from day 27 through day 548.
\end{abstract}

\section {Introduction}

Within days of the latest outburst of RS Ophiuchi on February 12.83
2006 \citep[taken as t=0]{2006IAUC.8671....2N}, a large number of
instruments began observing the nova.  These observations primarily
focused on the radio (e.g. O'Brien et al. 2006), infrared
(e.g. Evans et al. 2007a,b) and X-ray
\citep[e.g.][]{2006ApJ...652..629B}.  Whilst RS Oph was observed
optically by a large number of amateur instruments (see
AAVSO\footnote{http://www.aavso.org} light curve, reproduced by Starrfield in
these proceedings), few large
professional telescopes were employed in the optical.  Of the optical
data available many cover only a short time period and are concerned
mainly with rapid, short period variation \citep[see
  e.g.][]{2006IBVS.5733....1Z,2007MNRAS.379.1557W} or only cover
single epochs \citep[e.g.][]{2007ApJ...665L..63B_d}.  RS Oph represents
an ideal target for the fully robotic 2m Liverpool Telescope
\citep[LT;][]{2004SPIE.5489..679S_d} situated on La Palma, Canary
Islands.  In addition to its rapid response capability, the LT's
robotic nature is perfectly suited to monitor objects over extended
periods of time.  In this paper we present the LT broadband optical
photometry of the 2006 outburst of RS Oph.

\section {Observations}

LT broadband optical observations of RS Oph first became feasible on
March 11$^{th}$ 2006 (t=27.41 days), once its flux had
declined to $m_{r'}\sim7$.  From day 27 onwards the LT has monitored
RS Oph with a cadence (ignoring weather, seasonal or technical gaps)
of roughly 48 hours, through $z'$, $i'$, $r'$, $V$, $B$ and $u'$
filters and as of August 14$^{th}$ 2007 had obtained 73 epochs of
data.

\begin{figure}[t]
\includegraphics[clip=true,width=0.5\textwidth,angle=270]{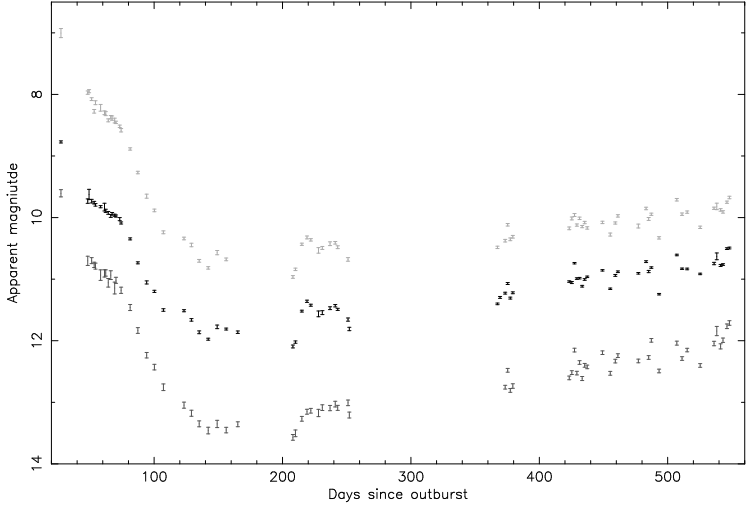}
\caption{LT optical light curve of RS Oph.  Light-grey (upper)
  Sloan-$r'$, black (centre) $V$ and Dark-grey (lower) $B$.  $z'$,
  $i'$ and $u'$ not shown (see text).}
\label{lightcurve}
\end{figure}

\begin{figure}[bh]
\includegraphics[clip=true,width=0.5\textwidth,angle=270]{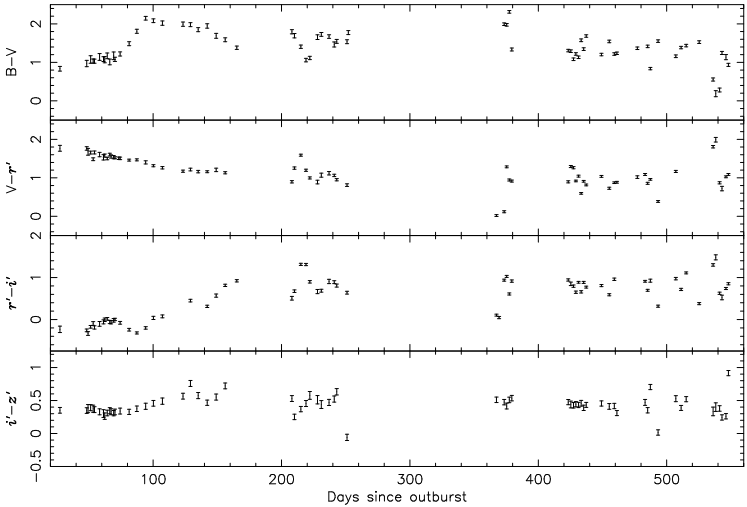}
\caption{$B$-$V$, $V$-$r'$, $r'$-$i'$, $i'$-$z'$ colour behaviour of
  RS Oph.}
\label{colour}
\end{figure}

The LT data have been reduced using standard profile fitting
photometry techniques provided by the IRAF package.  The photometry
has been independently verified using GAIA package.
Figure~\ref{lightcurve} shows the LT $r'$, $V$ and $B$ light curves of
RS Oph covering the first $\sim550$ days since outburst.  The $z'$ and
$i'$ light curves are not shown as they are similar in magnitude and
behaviour to the $r'$ curve.  Shown in Figure~\ref{colour} is the
colour behaviour of RS Oph since outburst over the same period. 


The ``break'' in the RS Oph light curve seen at t$\simeq$70 days (see
Figure~\ref{lightcurve}) is coincident (within the time resolution) to
a distinct minimum seen in the Si coronal line fluxes seen by Spitzer
\citep[see][]{2007ApJ...663L..29E_d}.  We also note that this break
coincides with a marked reddening in the B-V colour of RS Oph (see
Figure~\ref{colour}).  However, a break seen at 64 days in the 1-10
keV Swift light curve (see Bode et al., these proceedings) is not
apparent here.

These LT data identified a brightening in the optical luminosity of RS
Oph from approximately day 208 \citep{2006IAUC.8761....1B} to day 220.
Worters et al. (2007) found evidence of the resumption of
optical flickering by day 241, indicating a re-establishment of
accretion.  As can be seen in Figure~\ref{lightcurve}, following the
re-brightening, the luminosity of the system has steadily increased
and the light curve has shown a marked increase in variability
compared to the initial decline.  

LT monitoring of RS Oph is still on-going (as of summer 2008), although the cadence has
now been reduced to 72 hours.  The $u'$ data suffer from extremely low
signal-to-noise and are still being processed.  Also, we have employed
the experimental Ring Polarimeter
\citep[RINGO;][]{2006SPIE.6269E.179S_d} on the LT in an attempt to
measure any change in polarisation during the outburst.  To-date,
twelve epochs of RINGO observations have been obtained, however, these
data are yet to be fully reduced.

\acknowledgements

The Liverpool Telescope is operated on the island of La Palma by
Liverpool John Moores University in the Spanish Observatorio del Roque
de los Muchachos of the Instituto de Astrofisica de Canarias with
financial support from the UK Science and Technology Facilities
Council (STFC).

\end{document}